\documentclass[iop,onecolumn]{emulateapj}
\usepackage{graphicx}
\usepackage{subfigure}
\usepackage{amsmath}
\usepackage{amssymb}
\usepackage{comment}

\include{defns}
\def\gsim{\;\rlap{\lower 2.5pt
 \hbox{$\sim$}}\raise 1.5pt\hbox{$>$}\;}
\def\lsim{\;\rlap{\lower 2.5pt
   \hbox{$\sim$}}\raise 1.5pt\hbox{$<$}\;}
\defcitealias{liu09a}{Liu et al. (2009a)}
\defcitealias{liu09b}{Liu et al (2009b)}

\newcommand{\tr}[1]{\textrm{#1}}
\newcommand{\ee}[1]{\times10^{#1}}

\newcommand{\pp}[2]{\frac{\partial#1}{\partial#2}}

\shortauthors{Wiener, Zweibel, \& Oh}

\begin{document}
\bibliographystyle{mn2e}

\title{Cosmic Ray Heating of the Warm Ionized Medium}
\author{Joshua Wiener}
\affil{Department of Physics, University of California; Santa Barbara, CA 93106, USA}
\author{Ellen G. Zweibel}
\affil{Departments of Astronomy and Physics, and Center for Magnetic Self-Organization, University of Wisconsin-Madison, Madison, WI 53706, USA}
\author{S. Peng Oh}
\affil{Department of Physics, University of California; Santa Barbara, CA 93106, USA}

\begin{abstract}
  Observations of line ratios in the Milky Way's warm ionized medium (WIM) suggest that photoionization is not the only heating mechanism present. For the additional heating to explain the discrepancy it would have to have a weaker dependence on the gas density than the cooling rate, $\Lambda n_e^2$. \cite{reynolds99} suggested turbulent dissipation or magnetic field reconnection as possible heating sources. We investigate here the viability of MHD-wave mediated cosmic ray heating as a supplemental heating source. This heating rate depends on the gas density only through its linear dependence on the Alfv\'en speed, which goes as $n_e^{-1/2}$. We show that, scaled to appropriate values of cosmic ray energy density, cosmic ray heating can be significant. Furthermore, this heating is stable to perturbations. These results should also apply to warm ionized gas in other galaxies.
\end{abstract}

\section{Introduction}
Observations of [S II]/H$\alpha$ and [N II]/H$\alpha$ line intensity ratios in the WIM show a spatial variation with distance from the galactic midplane $|z|$ - larger line ratios are seen further from the disk (see \cite{reynolds99}, \cite{haffner09}). Such variation might be explained by variations in the ionization parameter $U$, the ratio of photon density to gas density. However, this would not explain the additional observation that the [S II]/[N II] ratio remains nearly constant with $|z|$. Under WIM conditions, variations in $U$ inevitably produces larger changes in Sulphur (which can be either in the form of SII or SIII) compared to Nitrogen (which almost always appears as NII), due to their different ionization potentials.

These observations may be explained by a spatial variation of the electron temperature $T_e$. An increase in $T_e$ with height above
the disk could explain the enhanced [S II]/H$\alpha$ and [N II]/H$\alpha$ ratios. Also, because [S II] and [N II] have nearly the same excitation energy, the [S II]/[N II] ratio is nearly independent of $T_e$. So the near constant [S II]/[N II] ratio may also be explained this way.

But how can this variation in $T_e$ be explained? 
If only photoionization heating is important, then this increase in $T_{\rm e}$ can potentially be accommodated by hardening of the spectrum away from the disk mid-plane. However, a hard spectrum is inconsistent with HeI $\lambda5876$ observations \citep{rand97,rand98}. On the other hand, if there were a secondary heating mechanism with a weaker dependence on electron density than the $n_{e}^{2}$ dependence of photoionization heating, such heating would dominate far from the disk, where densities are low, and we would see a variation in $T_e$ that could explain the observed line ratio variations.

Many such supplementary heating mechanisms have been proposed, such as photoelectric heating from dust grains (\cite{weingartner01}), magnetic reconnection (\cite{raymond92}), and turbulent dissipation (\cite{minter97}). We study here the possibility of cosmic ray heating. The process was outlined by \cite{wentzel71}, but not applied to the WIM (which had not been discovered at that time).
 If a cosmic ray population has a bulk velocity faster than the local Alfv\'en speed $v_{Ai}=B/\sqrt{4\pi \rho_i}$, magnetohydrodynamic Alfv\'en waves are generated and exhibit unstable growth
(\cite{wentzel68}, \cite{kulsrud69}). In a steady state,
these waves are damped by some other process(es), transferring energy to the gas. In this way a cosmic ray density gradient can indirectly heat the plasma. We will see that the resulting gas heating rate is proportional to $n_e^{-1/2}$, and is of the order required to explain the necessary temperature variations. Note that this process
is quite different from collisional cosmic ray heating (\cite{spitzer68}). A strength of this mechanism is that the heating rate depends on only a few parameters which are either observable or can be estimated. We first consider the nature of CR trapping in the WIM in \S\ref{equilibrium}, before considering the CR heating rate and its local stability properties in \S\ref{discussion}.

\section{Alfv\'en Wave Equilibrium}\label{equilibrium}
For cosmic ray heating to be in place we must ensure that Alfv\'en waves are present in the WIM with enough energy to scatter the cosmic rays. To do this we determine the wave damping, which in the WIM environment is due to ion-neutral friction and non-linear Landau damping. We then balance this damping with the cosmic-ray-induced wave growth to obtain an equilibrium condition. This condition determines the power spectrum in the waves. We can then derive a mean free path for the cosmic rays and determine if they are well-trapped. We could also use the equilibrium damping rate to determine the heating of the gas, but as we will see the heating rate depends only on the characteristics of the cosmic ray population, \emph{provided} they are well-trapped. 

\subsection{Ion-Neutral Damping in Nearly Ionized Gas}\label{ind}
We follow Appendix C of \cite{kulsrud69}. These authors assumed the gas is nearly neutral; we assume it is almost fully ionized.
We begin with the force equations for the neutral and charged components of the gas respectively. Assuming the transverse velocities of each component are in the form of an oscillator $v_i=A_ie^{i(kz-\omega t)}$, we have
\begin{equation}
 \rho_n\omega^2v_n=-i\nu_{ni}\omega\rho_n(v_n-v_i)
\end{equation}
\begin{equation}
 \rho_i\omega^2v_i=\rho_i\omega_k^2-i\nu_{in}\omega\rho_i(v_i-v_n)
\end{equation}

These equations are essentially those of two coupled, damped oscillators, one of which is driven with frequency $\omega_k=kv_{Ai}$, the natural Alfv\'en wave frequency. The neutral-ion collisions are treated as a drag force parameterized by the collision frequency $\nu_{ni}$. The movement of the ions by the Alfv\'en waves is impeded by the neutral particle population. This nudges the neutral component to follow behind the oscillating ions, removing energy from the Alfv\'en waves to do so.

We can rearrange these equations into convenient matrix form by using $\nu_{ni}/\nu_{in}=\rho_i/\rho_n$:
\begin{equation}
 (\omega^2-\omega_k^2+i\omega\nu_{ni})v_i-i\omega\nu_{ni}v_n=0
\end{equation}
\begin{equation}
 -i\omega\nu_{in}v_i+(\omega^2+i\omega\nu_{in})v_n=0
\end{equation}

Setting the determinant of this matrix to zero gives us the dispersion relation for $\omega$.
\begin{equation}
 \omega^3+i\nu_{ni}(1+\frac{\rho_n}{\rho_i})\omega^2-\omega_k^2\omega-i\nu_{ni}\omega_k^2=0
\end{equation}

We can solve this perturbatively in the two limits $\omega_k\ll\nu_{ni}$, $\omega_k\gg\nu_{ni}$. Let us further assume we are in the $\epsilon=\rho_n/\rho_i\ll1$ limit, to match the WIM\footnote{Note that our $\epsilon$ is the inverse of the one used in \cite{kulsrud69}}. Let's first consider the $\omega_k\ll\nu_{ni}$ case. To leading order in $\omega_k/\nu_{ni}$ we have
\begin{equation}
i\nu_{ni}(1+\epsilon)\omega_0^2-i\nu_{ni}\omega_k^2=0\qquad\Rightarrow \omega_0=\omega_k\sqrt{1+\epsilon}
\end{equation}
The first order equation is
\[
 \omega_0^3+2i\nu_{ni}(1+\epsilon)\omega_0\omega_1-\omega_k^2\omega_0=0
\]
\begin{equation}
\Rightarrow \omega_k^3\epsilon\sqrt{1+\epsilon}+2i\nu_{ni}(1+\epsilon)\omega_k\sqrt{1+\epsilon}\omega_1=0\qquad\Rightarrow \omega_1=-\frac{i\omega_k^2}{2\nu_{ni}}\frac{\epsilon}{1+\epsilon}
\end{equation}
So in this limit the waves are damped at a rate
\begin{equation}
 \Gamma_{\tr{in}}=\frac{i\omega_k^2}{2\nu_{ni}}\frac{\epsilon}{1+\epsilon}
\end{equation}

In the short wave limit, $\omega_k\gg\nu_{ni}$, which is the relevant limit for CR-generated waves, we have
\begin{equation}
 \omega_0^3-\omega_k^2\omega_0=0\\
\Rightarrow \omega_0=\omega_k
\end{equation}
\[
 3\omega_k^2\omega_1+i\nu_{ni}(1+\epsilon)\omega_k^2-\omega_k^2\omega_1-i\nu_{ni}\omega_k^2=0
\]
\begin{equation}\Rightarrow 2\omega_k^2\omega_1+i\nu_{ni}\epsilon\omega_k^2=0\qquad\Rightarrow \omega_1=-\frac{i\nu_{ni}}{2}\epsilon=-\frac{i\nu_{in}}{2}
\end{equation}
and so we have damping rate
\begin{equation}
 \Gamma_{\tr{in}}=-\frac{\nu_{in}}{2}
\end{equation}
Note that this is the same damping rate derived in \cite{kulsrud69}, even though we are in the opposite limit of a mostly ionized gas rather than a mostly neutral one.

The ion-neutral collision frequency is
\begin{equation}\label{nuin}
 \nu_{in}=\frac{m_n}{m_i+m_n}n_n\langle\sigma v\rangle=\frac{1}{2} n_n\langle\sigma v\rangle=\frac{1}{2}\epsilon n_i\langle\sigma v\rangle
\end{equation}
Here $n_n$ is the density of the neutral component and $\langle\sigma v\rangle$ is the average rate of exchange of velocity per particle for ion-neutral collisions. For temperatures around $10^4$ K, we have $\langle\sigma v\rangle\approx10^{-8}\tr{ cm}^3\tr{ s}^{-1}$ (see \cite{kulsrud71}, \cite{depontieu01}). We assume here that hydrogen is the dominant neutral species; up to 10\% of the hydrogen in the WIM is thought to be neutral (\cite{haffner09}); He should be mostly neutral, but has a lower collision rate (\cite{depontieu01}).
So in the short-wave, almost completely ionized limit, the ion-neutral damping rate is a function of only the density of the neutral component
\begin{equation}\label{inrate}
 \Gamma_{in}=-\frac{1}{4}n_n\langle\sigma v\rangle=-\frac{1}{4}\epsilon n_i\langle\sigma v\rangle.
\end{equation}

\subsection{Non-Linear Landau Damping}
In some circumstances, non-linear Landau damping may be comparable to or dominate over ion-neutral damping. Non-linear Landau damping occurs when ions ride along the envelope of a beat wave formed by two interfering Alfv\'en waves. Ions whose random motions are slightly slower than the speed of this envelope will take energy from the waves, damping them. Ions that are slightly faster will give energy to the waves, but for a thermal distribution we expect there to be more of the slower particles, and so the net effect is a wave damping. The strength of this damping depends on the strength of the waves as (\cite{kulsrud78})
\begin{equation}\label{nlldrate}
 \Gamma_\tr{NLLD}=-\sqrt{\frac{\pi}{8}}v_ik\left(\frac{\delta B}{B}\right)^2_k
\end{equation}
Here, $v_i=\sqrt{k_BT/m}$ is the thermal speed of the ions.

\subsection{Cosmic Ray Instability}
A cosmic ray traveling along a magnetic field line with speed $v$ and pitch angle cosine $\mu$ will interact with an Alfv\'en wave with parallel wave number $k_z$ under the resonance condition
\begin{equation}
 k_z=\frac{\sqrt{1-\mu^2}}{\mu r_L}=\frac{\Omega_0}{\gamma v\mu}
\end{equation}
Here, $r_L$ is the cosmic ray's relativistic gyroradius, $\Omega_0=eB_0/mc$ is its nonrelativistic gyrofrequency, and $\gamma$ is its Lorentz factor. In other words, a cosmic ray and an Alfv\'en wave are resonant if the wave's wavelength is roughly equal to the distance the cosmic ray travels along the B-field in one gyration.

\cite{wentzel68},
\cite{kulsrud69} showed that a population of cosmic rays whose bulk velocity is faster than the Alfv\'en speed will spur unstable growth in the waves. If we have such a distribution of CRs $f(\mathbf{x},\mathbf{p},t)$, the resulting growth rate can be written, in the wave frame, (\cite{skilling71}):
\begin{equation}\label{growth}
 \Gamma_{\tr{growth}}(k_z)=\frac{\pi^2m^2\Omega_0^2v_A}{2k_zB^2}\int d^3\mathbf{p}(1-\mu^2)v\pp{f}{\mu}\left[\delta\left(\mu p-\frac{m\Omega_0}{k_z}\right)+\delta\left(\mu p+\frac{m\Omega_0}{k_z}\right)\right]
\end{equation}
The above holds for Alfv\'en waves propagating nearly parallel to the background magnetic field $B\hat{z}$. From here on we drop the $z$ subscripts. The delta functions encode the resonance condition for CRs travelling in both directions.

We can rewrite this expression in terms of the cosmic ray gradient along the background magnetic field $\pp{f}{z}$. In the absence of any sources or sinks, the cosmic ray transport equation is
\begin{equation}\label{crt}
 \pp{f}{t}+\mu v\pp{f}{z}=\pp{}{\mu}\left[\frac{(1-\mu^2)}{2}\nu(\mu)\pp{f}{\mu}\right]
\end{equation}
The scattering frequency $\nu$ is related to the energy density $\mathcal{E}$ of resonant Alfv\'en waves (\cite{kulsrud69}:
\begin{equation}\label{nu}
 \nu(\mu)=\frac{2\pi^2\Omega}{B^2}k\mathcal{E}(k)=\frac{\pi}{4}\Omega\left(\frac{\delta B}{B}\right)^2,\qquad k=\frac{1}{|\mu|r_L}
\end{equation}
This collision frequency is expected to be very large compared to the cosmic ray dynamical timescale
 (a condition we must check later for consistency), so we can expand $f$ in inverse powers of $\nu$, $f=f_0+f_1+f_2+...$. To lowest order, eqn. (\ref{crt}) becomes
\begin{equation}
 0=\pp{}{\mu}\left[\frac{(1-\mu^2)}{2}\nu(\mu)\pp{f_0}{\mu}\right]\Rightarrow \pp{f_0}{\mu}=0
\end{equation}
To first order we have
\begin{equation}
 \mu v\pp{f_0}{z}=\pp{}{\mu}\left[\frac{(1-\mu^2)}{2}\nu(\mu)\pp{f_1}{\mu}\right]
\end{equation}
If we integrate both sides over $\mu$,
\begin{equation}
 \pp{f_1}{\mu}=-\frac{v}{\nu}\pp{f_0}{z}
\end{equation}
We can now eliminate $\pp{f}{\mu}$ from eqn \ref{growth}:
\begin{equation}
 \Gamma_{\tr{growth}}(k)=-\frac{\pi^2m^2\Omega_0^2v_A}{2kB^2}\int d^3\mathbf{p}(1-\mu^2)\frac{v^2}{\nu}\pp{f}{z}\left[\delta\left(\mu p-\frac{m\Omega_0}{k}\right)+\delta\left(\mu p+\frac{m\Omega_0}{k}\right)\right]
\end{equation}
Plugging in equation \eqref{nu} for $\nu$ we have
\begin{multline}
 \Gamma_{\tr{growth}}(k)=-\frac{2\pi m^2\Omega_0^2v_A}{k\Omega(\delta B)^2_k}\int d^3p(1-\mu^2)v^2\pp{f}{z}\left[\delta\left(\mu p-\frac{m\Omega_0}{k}\right)+\delta\left(\mu p+\frac{m\Omega_0}{k}\right)\right]\\
=-\frac{2\pi m\Omega_0v_A}{k(\delta B)^2_k}\int_0^\infty 2\pi p^2dp\int_{-1}^1d\mu(1-\mu^2)pv\pp{f}{z}\left[\delta\left(\mu p-\frac{m\Omega_0}{k}\right)+\delta\left(\mu p+\frac{m\Omega_0}{k}\right)\right]\nonumber
\end{multline}
Integrating the delta function over $\mu$ gives
\begin{equation}\label{growth1}
 \Gamma_{\tr{growth}}(k)=-\frac{8\pi^2 m\Omega_0v_A}{k(\delta B)^2_k}\int_{p_k}^\infty dp v\pp{f}{z}(p^2-p_k^2)
\end{equation}
where we have denoted $p_k=m\Omega_0/k$.

To make eqn. (\ref{growth1}) look a bit simpler, let us rewrite the integral in terms of a unitless factor of order unity $A(k)$:
\[
 \int_{p_k}^\infty dp vf(p)(p^2-p_k^2)\equiv\frac{1}{4\pi}cn_\tr{CR}A(k)
\]
\begin{equation}\label{Adef}
 A(k)=\frac{1}{n_\tr{CR}}\int_{p_k}^\infty dp\beta f(p)4\pi(p^2-p_k^2),\qquad 0\le A(k)\le1
\end{equation}
and let's define a CR length scale by
\begin{equation}\label{Ldef}
 -\pp{n_\tr{CR}}{z}\equiv\frac{n_\tr{CR}}{L_\tr{CR}}
\end{equation}
The growth rate is then\footnote{In principle the quantity $A(k)$ could vary in space, but we ignore this possibility here. Alternatively we could adjust our definition of $L_\tr{CR}$ to include this effect.}
\begin{equation}\label{growthrate}
 \boxed{\Gamma_\tr{growth}(k)=\frac{2\pi m \Omega_0 v_A c}{k(\delta B)^2}\frac{n_\tr{CR}}{L_{CR}}A(k)}
\end{equation}

Without specifying a cosmic ray distribution $f$ we cannot say anything about $A(k)$. As an example, consider a power law in momentum, $f(\mathbf{x},p,t)=C(\mathbf{x},t)p^{-\alpha}$, with some lower momentum cutoff $p_c$ and normalization:
\[
 n_{\tr{CR}}(\mathbf{x},t)=\int_{p_c}^\infty4\pi p^2f(\mathbf{x},p,t)dp=\frac{4\pi}{\alpha-3}Cp_c^{3-\alpha}
\]
\begin{equation}
\Rightarrow f(\mathbf{x},p,t)=n_{\tr{CR}}(\mathbf{x},t)\frac{\alpha-3}{4\pi p_c^3}\left(\frac{p}{p_c}\right)^{-\alpha}\Theta(p-p_c)
\end{equation}
Then by definition \eqref{Adef} we get
\[
 A(k)=\frac{\alpha-3}{4\pi p_c^3}\int_{p_k}^\infty dp\beta 4\pi(p^2-p_k^2)\left(\frac{p}{p_c}\right)^{-\alpha}\Theta(p-p_c)
\]
\[
=(\alpha-3)\int_{\tr{max}(x_k,1)}^\infty dx\beta(x^{2-\alpha}-x_k^2x^{-\alpha}),\qquad x\equiv\frac{p}{p_c}\qquad x_k\equiv\frac{p_k}{p_c}
\]
If we take the relativistic limit $\beta\approx1$ and denote $\tr{max}(x_k,1)=y_k$ this becomes
\[
 A(k)=(\alpha-3)\left[-\frac{y_k^{3-\alpha}}{3-\alpha}+\frac{x_k^2y_k^{1-\alpha}}{1-\alpha}\right]=y_k^{1-\alpha}\left[y_k^2-\frac{\alpha-3}{\alpha-1}x_k^2\right]
\]
or
\begin{equation}\label{ak}
  A(k)=
\begin{cases}
 \frac{2}{\alpha-1}\left(\frac{k}{k_c}\right)^{\alpha-3} & k<k_c\\
 \left[1-\frac{\alpha-3}{\alpha-1}\left(\frac{k_c}{k}\right)^2\right] & k>k_c
\end{cases}
\end{equation}
where $k_c=m\Omega_0/p_c$ is determined from the lower momentum cutoff of the spectrum.

\subsection{Equilibrium Power Spectrum}
Now that we have the total damping and growth rates of the Alfv\'en waves we can enforce an equilibrium condition
\begin{equation}\label{gammagamma}
 \Gamma_\tr{growth}+\Gamma_\tr{in}+\Gamma_\tr{NLLD}=0
\end{equation}
Inserting our expressions \eqref{nlldrate}, and \eqref{growthrate} into eqn. (\ref{gammagamma}),
\[
 \frac{2\pi m \Omega_0 v_A c}{k(\delta B)^2}\frac{n_\tr{CR}}{L_{CR}}A(k)-\Gamma_{in}-\sqrt{\frac{\pi}{8}}v_ik\left(\frac{\delta B}{B}\right)^2_k=0
\]
or, rearranging terms,
\[
 \sqrt{\frac{\pi}{8}}v_ik^2\mathcal{X}^2+\Gamma_{in} k\mathcal{X}-\frac{2\pi m \Omega_0 v_A c}{B^2}\frac{n_\tr{CR}}{L_{CR}}A(k)=0,\qquad\mathcal{X}\equiv\left(\frac{\delta B}{B}\right)^2_k
\]
We solve analytically for ${\mathcal{X}}$
\begin{equation}\label{X}
 \left(\frac{\delta B}{B}\right)^2_k=\sqrt{\frac{2}{\pi}}\frac{\Gamma_{in}}{kv_i}[-1+\sqrt{1+\mathcal{R}}],\qquad \mathcal{R}\equiv\sqrt{\frac{\pi}{2}}\frac{c}{v_A}\frac{r_i}{L_{CR}}\frac{n_{CR}}{n_i}\frac{\Omega_0^2}{\Gamma_{in}^2}A(k),
\end{equation}
where $r_i\equiv v_i/\Omega_0$ is the thermal ion gyroradius.

It is informative to determine the relative importance of each damping mechanism. We can do this by looking at the quantity $\mathcal{R}$. If ${\mathcal{R}}$ is small, the linear term (ion-neutral damping) dominates and
\begin{equation}\label{smallR}
\left(\frac{\delta B}{B}\right)_k^2 \approx\frac{1}{2}\frac{c}{v_A}\frac{n_{CR}}{n_i}\frac{\Omega_0A(k)}{kL_{CR}\Gamma_{in}},\qquad {\mathcal{R}}\ll 1.
\end{equation}
 while if ${\mathcal{R}}$ is large, non-linear Landau damping dominates, and
\begin{equation}\label{largeR}
\left(\frac{\delta B}{B}\right)_k^2 \approx\left(\sqrt{\frac{2}{\pi}}\frac{c}{v_A}\frac{n_{CR}}{n_i}\frac{A(k)}{k^2r_iL_{CR}}\right)^{1/2}, \qquad {\mathcal{R}}\gg 1.
\end{equation}
The transition between these two limiting cases occurs at $\mathcal{R}=1$, or
\begin{equation}\label{transition}
A(k)\approx 2.8 \ee{-4}\left(\frac{v_i}{10^6\tr{ cm/s}}\right)^{-1}\left(\frac{n_i}{.01\tr{ cm}^{-3}}\right)^{5/2}\left(\frac{n_\tr{CR}}{10^{-9}\tr{ cm}^{-3}}\right)^{-1}\left(\frac{L_\tr{CR}}{\tr{kpc}}\right)\left(\frac{\epsilon}{.05}\right)^2
\end{equation}
where we have used eqn. (\ref{inrate}) and taken
$\langle\sigma v\rangle=10^{-8}\tr{ cm}^3\tr{ s}^{-1}$. We introduce here a set of convenient fiducial values that we will use throughout this paper. For the power law spectrum with $\alpha=4.7$ this gives us a transition wave number $k^*$ of
\begin{equation}
 k^*\approx .012k_c=.012\frac{eB}{p_cc}=3.6\ee{-15}\tr{ cm}^{-1}\left(\frac{B}{\mu\tr{G}}\right)\left(\frac{p_cc}{\tr{GeV}}\right)^{-1}
\end{equation}
at the fiducial values in \eqref{transition}\footnote{We note here that the value of $k^*$ depends heavily on these quantities, particularly the gas density $n_i$. In fact, for some values there is no region of $k$-space where non-linear Landau damping dominates.}. For $k\ll k^*$, ion-neutral damping dominates and the wave power is given by \eqref{smallR}. For $k\gg k^*$, non-linear Landau damping is dominant and the wave power is \eqref{largeR}.

\begin{figure}[h!]
\begin{center}
\includegraphics[height=50mm,trim=0cm 1cm 0cm -1cm]{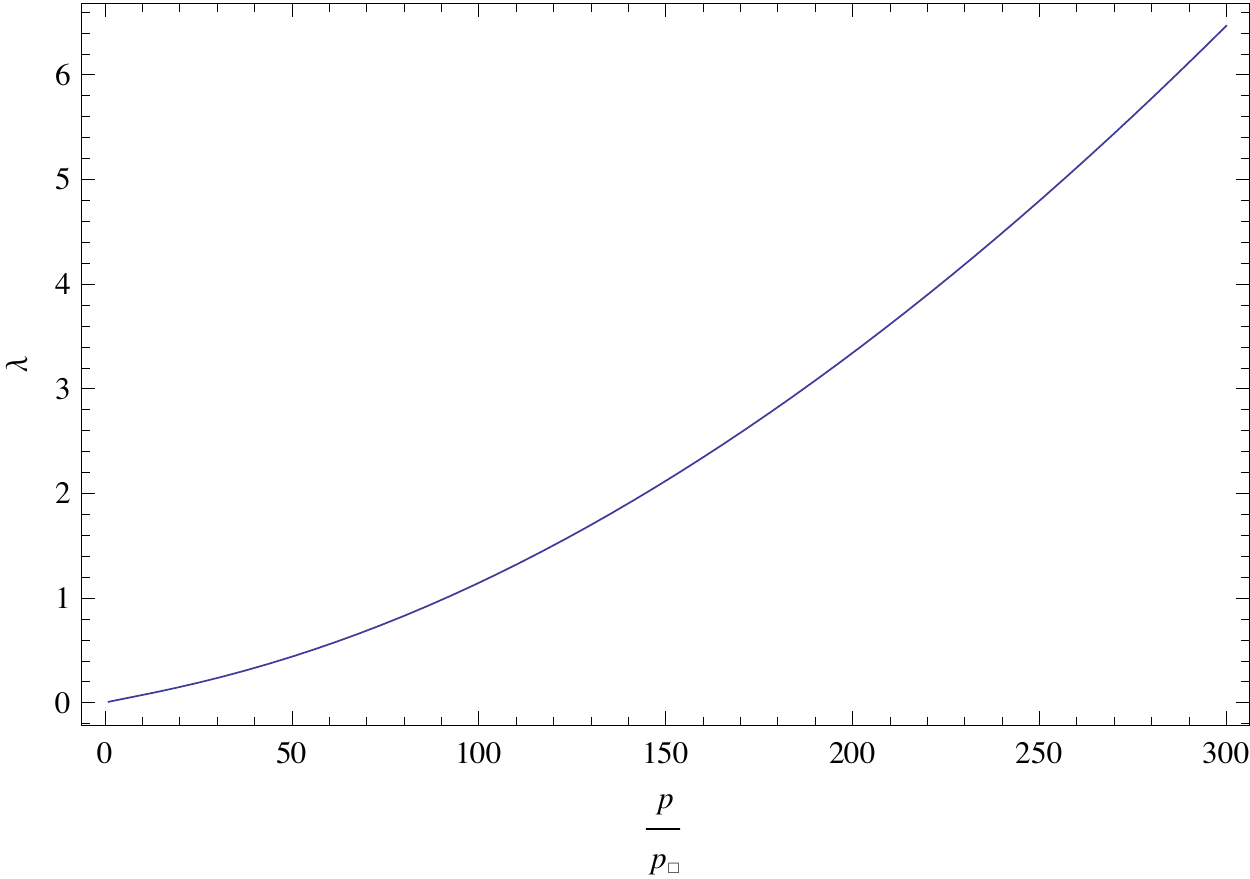}
\end{center}
\caption{Cosmic ray mean free path to scattering by self generated turbulence, calculated from eqn. (\ref{lambda2}) with the parameters set equal to the fiducial values. The mean free path is given in pc and the cosmic ray momentum is given in terms of the cutoff momentum $p_c$.}
\label{fig:lambda}
\end{figure}

We are now in a position to check whether the cosmic rays are self-trapped, i.e. whether their mean free path to scattering by self generated turbulence is small compared to their scale height. This is
a necessary condition for applying the heating theory derived in \S\S\ref{sec:heatingrate} and \ref{discussion}. The mean free path $\lambda$ is related to the scattering frequency given in eqn.
(\ref{nu}) by
\begin{equation}\label{lambda1}
\lambda = \frac{v}{\nu}.
\end{equation}
Using eqns. (\ref{nu}) and (\ref{X}) in eqn. (\ref{lambda1}) yields a relatively compact expression for $\lambda$
\begin{equation}\label{lambda2}
\lambda=\sqrt{\frac{8}{\pi}}\frac{v_i}{\Gamma_{in}}\left(-1+\sqrt{1+\mathcal{R}}\right)^{-1}.
\end{equation}
The mean free path in pc given by eqn. \eqref{lambda2} is plotted as a function of $p$ in units of the cutoff momentum $p_c$ in Figure \ref{fig:lambda}.
The figure spans the transition from Landau damping dominated at low momentum to ion-neutral friction dominated at high momentum. It appears from Figure \ref{fig:lambda} that cosmic rays of energy even several hundred times the cutoff energy are quite well trapped ($\lambda\sim$ several pc). Most of the cosmic ray energy lies in the trans-relativistic $\sim$GeV regime. In the ISM, the spectra turns over at $p_{\rm c} \sim 10$ MeV due to Coulomb cooling. Thus, the regime of interest is $p/p_{\rm c} \sim 100$. At the upper end of the trapped range, where eqn. (\ref{smallR}) holds, eqn. \eqref{lambda2} can be written
in the form
\begin{equation}\label{lambdain}
\frac{\lambda}{L_{CR}}\approx\frac{8}{\pi}\frac{\Gamma_{in}}{\Omega_0}\frac{v_A}{c}\frac{n_i}{n_{CR}A(k)}\rightarrow 4.0\times 10^{-7}\left(\frac{p}{p_c}\right)^{1.7},
\end{equation}
where in the last expression we have used eqn. (\ref{ak}) and set all parameters to their fiducial values.

Figure \ref{fig:lambda} can also be used to check that the waves are small amplitude and well described by linear theory. From eqns. (\ref{nu}) and (\ref{lambda1}) was can see that $(\delta B/B)_k^2\sim
r_L/\lambda$. Cosmic rays of energy a few hundred GeV and less have gyroradii of order 10s of AU or less, showing that $\delta B/B\ll 1$ even at low momenta where the  mean free path is short.

\subsection{Heating Rate}\label{sec:heatingrate}
We can now determine the heating rate of the WIM due to the dissipation of Alfv\'en waves created by cosmic ray streaming. To do this we want to integrate the time-derivative of $\mathcal{E}(k)=\delta B_k^2/8\pi k$ over all wave numbers $k$. We know that the time-dependence of $\delta B$ in an Alfv\'en wave is
\begin{equation}
 \delta B\propto e^{-i\omega t}
\end{equation}
\[
 \omega=\omega_R+i\Gamma_\tr{damp}
\]
and so
\begin{equation}
 \mathcal{E}(k)=\frac{\delta B_k^2}{8\pi k}\propto e^{2\Gamma_\tr{damp}t}
\Rightarrow \pp{\mathcal{E}(k)}{t}=2\Gamma_\tr{damp}\mathcal{E}(k)
\end{equation}
(note that $\Gamma_\tr{damp} < 0$).

Let us remove the assumption of a power law spectrum and a relativistic limit and go back to any general distribution $f(\mathbf{x},p,t)$. Let us also remove any assumptions about damping mechanisms, and only assume we have equilibrium for the Alfv\'en waves. Then from equation \eqref{growthrate} and $\Gamma_\tr{growth}=\Gamma_\tr{damp}$,
\begin{equation}
 H=\int_0^\infty dk2\Gamma_\tr{damp}(k)\frac{\delta B_k^2}{8\pi k}=\int_0^\infty dk\Gamma_\tr{damp}(k)\frac{m\Omega_0 v_A c}{2\Gamma_\tr{damp}(k)k^2}\frac{n_\tr{CR}}{L_{CR}}A(k)
\end{equation}
\begin{equation}
 H=-\int_0^\infty dk\frac{m\Omega_0 v_A c}{2k^2}\pp{}{z}\left[\int_{p_k}^\infty dp\beta f(p)4\pi(p^2-p_k^2)\right]
\end{equation}
In the last step we have rewritten $A(k)$ and $L_\tr{CR}$ in terms of their original definitions \eqref{Adef} and \eqref{Ldef}. We reformulate this double integral with a change of variable from $k$ to $p_k$, $dp_k=-m\Omega_0/k^2dk$. Let us also write $\mathbf{v_A}=v_A\mathbf{n}$ such that
\begin{equation}
 H=-\frac{1}{2}\mathbf{v_A}\cdot\nabla\left[\int_0^\infty dp_k\int_{p_k}^\infty dp4\pi f(p)v(p)(p^2-p_k^2)\right]
\end{equation}
Finally, let's exchange the order of the integrals by recognizing that the double integral is over all $(p,p_k)$ under the constraint $0\le p_k\le p$.
\[
 H=-\frac{1}{2}\mathbf{v_A}\cdot\nabla\left[\int_0^\infty dp\int_0^p dp_k4\pi f(p)v(p)(p^2-p_k^2)\right]
\]
\begin{equation}
H=-\frac{1}{2}\mathbf{v_A}\cdot\nabla\left[\int_0^\infty dp 4\pi p^3 f(p)v(p)-\frac{1}{3}\int_0^\infty dp 4\pi p^3f(p)v(p)\right]
\end{equation}
These integrals are now very simple - they correspond to the total cosmic ray pressure
\begin{equation}
 P_\tr{CR}=\frac{1}{3}\int_0^\infty dp4\pi p^2f(p)v(p)
\end{equation}
We therefore obtain the very simple expression for the cosmic ray heating:
\begin{equation}\label{crheat}
\boxed{H=-\mathbf{v_A}\cdot\nabla P_\tr{CR}}
\end{equation}
in agreement with \cite{wentzel71}.

The heating rate is simply the cosmic ray pressure gradient times the Alfv\'en speed. Even without the above calculation we know this \emph{must} be the solution, since we require an equilibrium for the waves and \eqref{crheat} is always the rate at which cosmic rays give energy to the Alfv\'en waves regardless of damping (\cite{mckenzie81}). So, as hinted at in section \ref{equilibrium}, we require \emph{only} that the Alfv\'en waves are in equilibrium and the cosmic rays are well-trapped to know that the cosmic ray heating is \eqref{crheat}.

\section{Application to our Galaxy}
\label{discussion}
\subsection{Observations}\label{sec:observations}
Let us carry through the dependence of $v_A$ on the magnetic field and ion density, and pick some representative values. To write the cosmic ray pressure in terms of the energy density, we use $P_\tr{CR}=0.45E_\tr{CR}$ from \cite{ferriere01}. Then we can estimate the heating rate in the WIM:
\begin{equation}\label{heat0}
 H\approx 5.2\times10^{-28} \frac{\rm{erg}}{\rm{cm}^3\ \rm{s}}\frac{E_\tr{CR}}{\rm{eV cm}^{-3}}\frac{B}{\mu\rm{G}}\left(\frac{L}{\rm{kpc}}\right)^{-1}\left(\frac{n_i}{10^{-2}\ \rm{cm}^{-3}}\right)^{-1/2}
\end{equation}
If we assume the cosmic ray energy density and magnetic energy density fall off with the same scale height $L$, we can determine the dependence of this heating rate on height $z$ from the galactic plane
\begin{equation}
 B(z)=B_0e^{-|z|/2L},\qquad E_\tr{CR}(z)=E_{\tr{CR},0}e^{-|z|/L}
\end{equation}
\begin{equation}\label{heat}
  H(z)\approx 5.2\times10^{-28} \frac{\rm{erg}}{\rm{cm}^3\ \rm{s}}\frac{E_{\tr{CR},0}}{\rm{eV cm}^{-3}}\frac{B_0}{\mu\rm{G}}\left(\frac{L}{\rm{kpc}}\right)^{-1}\left(\frac{n_i}{10^{-2}\ \rm{cm}^{-3}}\right)^{-1/2}e^{-3|z|/2L}
\end{equation}

Let's compare this to the heating that would be necessary to explain the inferred temperature profile $T(z)$. Following the prescription in \cite{reynolds99} we find $T(z)$ in our model by solving a heating-cooling balance equation
\begin{equation}\label{balance}
 G_0n_e^2+G_3n_e^{-1/2}=\Lambda n_e^2
\end{equation}
where each term represents, from left to right, photoionization heating, cosmic ray heating, and the cooling rate. The temperature dependence of the electron density $n_e$, the cooling function $\Lambda$, and the photoionization heating $G_0$ are
\begin{equation}
 n_e(|z|)=0.125T_4^{0.45}f^{-0.5}e^{-|z|/1\tr{kpc}} \, {\rm cm^{-3}}
\end{equation}
\begin{equation}
 \Lambda=3.0\ee{-24}T_4^{1.9}\tr{ erg cm}^3\tr{ s}^{-1}
\end{equation}
\begin{equation}
 G_0=1.2\ee{-24}T_4^{-0.8}\tr{ erg cm}^3\tr{ s}^{-1}
\end{equation}
$T_4$ denotes the temperature in units of $10^4\tr{ K}$, and $f$ is a filling fraction describing the amount of ionized Hydrogen, which we will set to $f(z)=\tr{Min}[0.1e^{|z|/750\tr{pc}},1]$. Note that our assumptions imply that $\nabla P_{\rm c} < \rho g$ at all $z$, and is consistent with hydrostatic balance. 

\begin{figure}
\includegraphics[width=14cm,trim=-2cm 0cm 0cm 0cm]{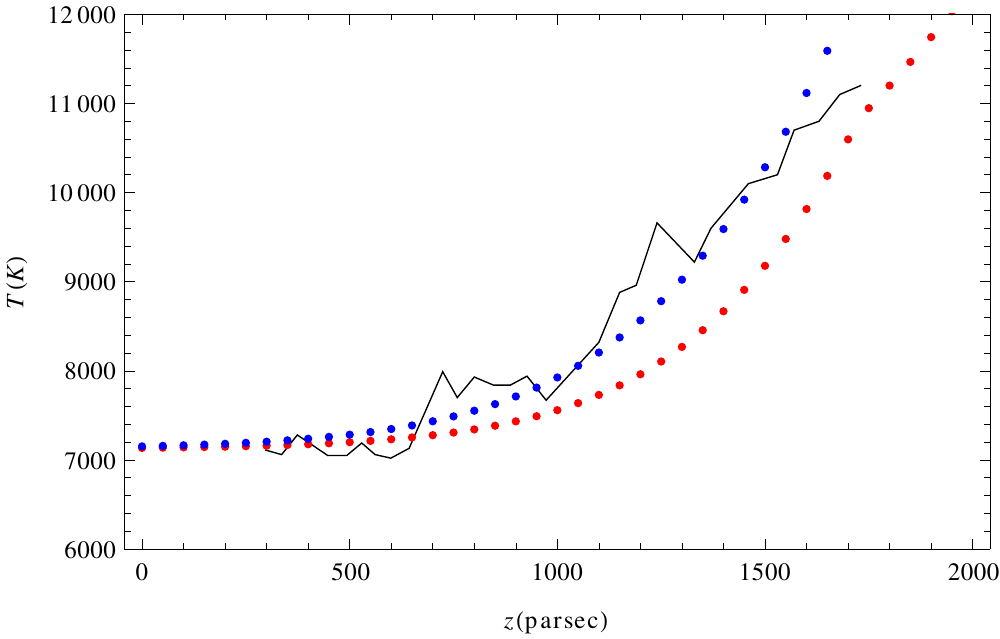}
\caption{Temperature versus height for the WIM. The black line is derived from observations of line ratios in the Perseus spiral arm. The blue points show the fit solution using the above parameters. The red points show the profile using parameters based on observations.}
\label{plot1}
\end{figure}

We then solve eqn \ref{balance} to obtain the model profile $T(z)$ and compare it to the profile infered from line ratio observations. We adjust $G_3$ to fit the model curve to the data, and we find, for $L=2\tr{ kpc}$:
\begin{equation}
 G_{3,\tr{fit}}\approx 1.2\ee{-27}\frac{\rm{erg}}{\rm{cm}^{9/2}\ \rm{s}}e^{-3|z|/4000\tr{ pc}}
\end{equation}
\begin{equation}
 \Rightarrow H_\tr{fit}=G_{3,\tr{fit}}n_e^{-1/2}=1.2\ee{-26}\frac{\rm{erg}}{\rm{cm}^3\ \rm{s}}\left(\frac{n_i}{10^{-2}\ \rm{cm}^{-3}}\right)^{-1/2}e^{-3|z|/4000\tr{ pc}}
\end{equation}
See figure \ref{plot1} (blue points) for this fit.

Comparing this to equation \eqref{heat} with $L_\tr{CR}=2$ kpc, we see that cosmic ray heating is sufficient to explain the observed line ratios if the magnetic field and CR energy density normalizations are high enough:
\begin{equation}\label{requirement}
 \frac{E_{\tr{CR},0}}{\rm{eV cm}^{-3}}\frac{B_0}{\mu\rm{G}}\approx46
\end{equation}

Is this the case? The magnetic field and CR energy density in the solar neighborhood are about $B_0=5\ \mu\tr{G}$ and $E_\tr{CR,0}=1.8\tr{ ev cm}^{-3}$ (\cite{ferriere01}). \cite{beuermann85} showed that synchrotron emissivity in the galactic spiral arms is about 4 times greater than in the interarm regions. Synchrotron emissivity depends on the field and CR density as (\cite{pfrommer04})
\begin{equation}
 j_\nu\propto E_\tr{CR}B^{\alpha/2+1}\propto B^{\alpha/2+3}
\end{equation}
for a power law CR density. We utilize our equipartition assumption in the last step. A spiral-arm enhancement of $j_\nu$ by a factor of 4 therefore implies the $B$-field increases by a factor of about 1.4 and the CR density increases by about 1.9 (for $\alpha=4.7$). We might then expect the product of the $B$-field and the CR density in the Perseus arm to be about
\begin{equation}\label{estimate}
 E_{\tr{CR,0}}B_0\approx(1.8\tr{ eV cm}^{-3})(5\ \mu\tr{G})*1.4*1.9\approx 24\ \mu\tr{G eV cm}^{-3}
\end{equation}
This falls short of the requirement from equation \eqref{requirement} but still comes close to reproducing the inferred temperature profile, as shown in the red points in figure \ref{plot1}. We also have not incorporated the orientation of the magnetic field. We have assumed a vertical field, but the actual field in the WIM may be much more random. This might be accounted for with an effective efficiency parameter.

One complication these estimates do not fully take into account is the multiphase nature of the ISM: the WIM has a low ($\sim 20\%$) filling factor, and above the disk is interspersed with hot diffuse coronal gas. This can cause local variations in Alfven speed, and thus cosmic ray pressure. As long as $B$ increases more slowly than $n_{e}^{1/2}$ (note that in the cooler diffuse ISM sampled by HI lines, $B$ as measured by Zeeman splitting does not scale with density \citep{crutcher10}), $v_{A}$ will be reduced in the WIM relative to coronal gas, and thus $P_{c}$ is higher. Thus, the relevant length scale for cosmic ray pressure gradients could be the cloud size, rather than the global scale height we have adopted; this leads to larger heating rates\footnote{It also implies most heating occurs when the cosmic rays exit the cloud, when $v_{\rm A} \cdot \nabla P_{\rm c} < 0$.}. On the other hand, if clouds are sufficiently small ($L_{\rm cloud} \lsim \lambda \sim 10$pc), then the CRs smooth over these inhomogeneities and the global gradients are appropriate. H$\alpha$ observations and photoionization modeling indicate that the WIM is likely to have both a smooth and a clumpy component which fluctuates on a wide range of length scales, but there is no consensus picture \citep{haffner09}. We regard these issues as beyond the scope of this paper, but such considerations are illustrative of possible variations in the cosmic rate heating rate. 

\subsection{Stability}\label{sec:stability}
We must check that the heating from the cosmic ray pressure gradient is stable under perturbations. If the heating increases compared to the cooling for a small element of gas perturbed to a higher temperature, the heating is unstable and we would get thermal runaway. To determine if this is the case, we need the change in the heating rate for a given perturbation.
\begin{equation}
\delta(\Lambda n^2-H)=\delta(L_0n^2T^{1.9}-H_0n^2T^{-0.8}-H_3BE_\tr{CR}n^{-1/2})
\end{equation}
In terms of perturbed quantities this becomes
\begin{multline}
 =L_0n^2T^2\left(2\frac{\delta n}{n}+1.9\frac{\delta T}{T}\right)-H_0n^2T^{-0.8}\left(2\frac{\delta n}{n}-0.8\frac{\delta T}{T}\right)
-H_3BE_\tr{CR}n^{-1/2}\left(\frac{\delta B}{B}+\frac{\delta E_\tr{CR}}{E_\tr{CR}}-\frac{1}{2}\frac{\delta n}{n}\right)
\end{multline}
Making use of the fact that the initial state was in thermal equilibrium, $L_0n^2T^2=H_0n^2T^{-0.8}+H_3BE_\tr{CR}n^{-1/2}$, we can write this most generally as
\begin{multline}\label{stability}
 \delta(\Lambda n^2-H)=\frac{\delta T}{T}\left[1.9L_0n^2T^{1.9}+0.8H_0n^2T^{-0.8}\right]\\
+\frac{\delta n}{n}\left[2L_0n^2T^{1.9}-2H_0n^2T^{-0.8}-\frac{1}{2}H_3BE_\tr{CR}n^{-1/2}\right]-H_3BE_\tr{CR}n^{-1/2}\left(\frac{\delta B}{B}+\frac{\delta E_\tr{CR}}{E_\tr{CR}}\right)\\
=\frac{\delta T}{T}\left[1.9L_0n^2T^{1.9}+0.8H_0n^2T^{-0.8}\right]-H_3BE_\tr{CR}n^{-1/2}\left(\frac{3}{2}\frac{\delta n}{n}+\frac{\delta B}{B}+\frac{\delta E_\tr{CR}}{E_\tr{CR}}\right)
\end{multline}
Without specifying the perturbation this is as far as we can go. Once we relate the perturbed quantities with $\delta T$, we can determine whether the heating is stable or unstable.

Let us consider an isobaric perturbation perpendicular to the magnetic field lines. The total pressure remains constant
\begin{equation}
 P_\tr{tot}=P_g+P_B+P_\tr{CR}=const.
\end{equation}
The above terms represent the gas pressure, magnetic pressure, and CR pressure respectively. The field lines are compressed along with the gas, so
\begin{equation}\label{bb}
 \frac{\delta B}{B}=\frac{\delta n}{n}
\end{equation}
Let us further assume that the cosmic rays respond adiabatically:
\begin{equation}\label{crcr}
 \frac{\delta E_\tr{CR}}{E_\tr{CR}}=\frac{\delta P_\tr{CR}}{P_\tr{CR}}=\gamma_\tr{CR}\frac{\delta n}{n}
\end{equation}
Then we can use the constant pressure condition to relate $\delta n$ and $\delta T$. From $P_g\propto nT$ and $P_B\propto B^2$ we get
\[
 \delta P_\tr{tot}=\delta P_g+\delta P_B+\delta P_\tr{CR}=0
\]
\[
 P_g\left(\frac{\delta n}{n}+\frac{\delta T}{T}\right)+2P_B\frac{\delta B}{B}+\gamma_\tr{CR}P_\tr{CR}\frac{\delta n}{n}=0
\]
\begin{equation}
 \frac{\delta n}{n}=-\frac{P_g}{P_g+2P_B+\gamma_\tr{CR}P_\tr{CR}}\frac{\delta T}{T}
\end{equation}

Putting this all together into equation \eqref{stability} gives
\begin{multline}
 \delta(\Lambda n^2-H)=\frac{\delta T}{T}\left[1.9L_0n^2T^{1.9}+0.8H_0n^2T^{-0.8}-\right.
\left.H_3BE_\tr{CR}n^{-1/2}\left(\frac{3}{2}-\gamma_\tr{CR}\right)\left(\frac{P_g}{P_g+2P_B+\gamma_\tr{CR}P_\tr{CR}}\right)\right]
\end{multline}
If the term in the brackets is positive, a small increase in temperature causes the change in cooling to outweigh the change in heating and the perturbation is stable. If the term in brackets is negative, it is unstable. But note that both parenthesised factors in the third term must each be less than one. Also, by the thermal equilibrium condition, $H_3BE_\tr{CR}n^{-1/2}$ must be less than $L_0n^2T^{1.9}$. The first term must therefore be of higher magnitude than the third term, and so the expression in the brackets is positive and the heating is stable.

This is perhaps easier seen if we denote the total cooling by $C$, and the fraction of the total heating due to photoelectric heating by $0\le x\le 1$. Then,
\begin{equation}
 \delta(\Lambda n^2-H)=C\frac{\delta T}{T}\left[1.9+0.8x-(1-x)\left(\frac{3}{2}-\gamma_\tr{CR}\right)\left(\frac{P_g}{P_g+2P_B+\gamma_\tr{CR}P_\tr{CR}}\right)\right]
\end{equation}

Now let us consider an acoustic perturbation. Equations \eqref{stability},\eqref{bb},and \eqref{crcr} still hold. Pressure is no longer fixed, but the gas responds almost adiabatically. As such
\begin{equation}
 \frac{P_g}{\rho^{\gamma_g}}=const.\rightarrow \delta\left(\frac{P_g}{\rho^{\gamma_g}}\right)=0
\end{equation}
\begin{equation}
\Rightarrow\frac{P_g}{\rho^{\gamma_g}}\left(\frac{\delta P_g}{P_g}-\gamma_g\frac{\delta \rho}{\rho}\right)=\frac{P_g}{\rho^{\gamma_g}}\left(\frac{\delta T}{T}+(1-\gamma_g)\frac{\delta n}{n}\right)=0\Rightarrow\frac{\delta n}{n}=\frac{1}{\gamma_g-1}\frac{\delta T}{T}
\end{equation}
We therefore have
\begin{equation}
 \delta(\Lambda n^2-H)=\frac{\delta T}{T}\left[1.9L_0n^2T^{1.9}+0.8H_0n^2T^{-0.8}-H_3BE_\tr{CR}n^{-1/2}\frac{3/2-\gamma_\tr{CR}}{\gamma_g-1}\right]
\end{equation}
or, in terms of $C$ and $x$,
\begin{equation}
 \delta(\Lambda n^2-H)=C\frac{\delta T}{T}\left[1.9+0.8x-(1-x)\frac{3/2-\gamma_\tr{CR}}{\gamma_g-1}\right]
\end{equation}
By a similar argument as before, this is also stable.

\section{Summary and Conclusions}

The gaseous disk of the Milky Way has a warm ionized component (WIM) with scale height several times that of the predominantly neutral component. The magnetic field and cosmic ray components have  similar thickness. Thick layers of warm ionized gas, and extended nonthermal emission, are seen in other galaxies as well \citep{haffner09}.

It is widely accepted that starlight photoionizes and heats the WIM.  Nevertheless, there is evidence for a supplemental heating mechanism. Detailed reconstruction of the WIM vertical temperature profile in the region of the Perseus spiral arm shows an increase in temperature with height that cannot be explained by radiative heating alone (\cite{reynolds99}). These authors showed that these observations can be explained by an additional heating mechanism with a weaker density dependence than the $n^2$ dependence of radiative heating. Heating by magnetic reconnection (\cite{raymond92}), dissipation of turbulence (\cite{minter97}), and photoelectric heating by dust (\cite{weingartner01}) have all been invoked. All three are feasible on energetic grounds, but the rates of the first two, in particular, depend on many unknown factors and are quite uncertain.

In this paper, we estimated the heating rate due to dissipation of waves excited by streaming cosmic rays. When the cosmic rays are well scattered by this self-generated turbulence, the heating rate depends only on the cosmic ray pressure gradient projected along the local magnetic field direction and the magnitude of the Alfv\'en speed (eqn. \ref{crheat}). Cosmic ray heating of the interstellar medium was discussed in general in \cite{wentzel71} and is included in models of supernova driven shock waves (\cite{volk84}), cosmic ray driven galactic winds (\cite{breitschwerdt91}, \cite{everett08}), and diffuse interstellar clouds (\cite{everett11}) but up to now does not appear to have been considered for the WIM. In \S 2 we showed that when wave excitation by streaming is balanced against nonlinear Landau damping and ion-neutral friction, the resulting wave amplitude, while small enough to allow treating the waves in the small amplitude approximation, is large enough to scatter the majority of cosmic rays many times over one pressure scale height (Figure 1). This justifies the frequent scattering limit we used in \S\ref{sec:observations}  to estimate the cosmic ray heating rate for the WIM (eqns \ref{heat0} and \ref{heat}) and show that adding it to the thermal equilibrium model of the WIM for Perseus Arm conditions produces a reasonably good fit to the observations (Figure 2).  Although the heating rate coefficient is about a factor of 2 too small (eqn. \ref{estimate}), the height dependence - which follows from the height dependence of the magnetic field, gas density, and cosmic ray pressure - leads to a temperature vs height relation of the correct shape. We regard this, and matching the inferred size of the supplemental heating rate to within a factor of two -  as confirmation that cosmic ray heating is a viable supplementary heat source for the WIM. Cosmic ray heating also seems to be a thermally stable mechanism (\ref{sec:stability}), at least under the assumptions we considered.

The results in this paper should be generally applicable to warm ionized gas in other galaxies. In cases where synchrotron emission is detected or other estimates of the cosmic ray and magnetic field energy densities are available, it should be possible to estimate the magnitude of cosmic ray heating. It is important that the gas be diffuse and that the ionization fraction be high; in weakly ionized clouds, for example, ion-neutral friction is so strong that the cosmic rays are not well coupled to the medium (\cite{everett11}). And, as long as the cosmic rays are well scattered, their pressure gradient along the ambient magnetic field exerts a force which may be important in determining the scale height of the gas and in driving an outflow even when the thermal speed of the gas is well below what is needed for escape.
\acknowledgements

We are happy to acknowledge useful discussions with J.S. Gallagher, M. Haffner and R. Reynolds and support from NSF Grants PHY0821899 and AST0907837 to the University of Wisconsin. JW and SPO acknowledge support from NASA grant NNX12AG73G to UCSB.  
\bibliography{master_references}

\end{document}